 \makeatother

\documentclass[aps,pre,floats,twocolumn,superscriptaddress,%
  amssymb,amsmath,showpacs]{revtex4}
\usepackage{graphicx}
\usepackage{chemarr}
\usepackage{epstopdf}
\usepackage[hypertex]{hyperref}
\usepackage{color}
\usepackage{subfigure}
\usepackage{hyperref}
\usepackage{pifont}
\begin{document}
\title{Impact of Charge Variation on the Encapsulation of Nanoparticles by Virus Coat Proteins}
\author{Hsiang-Ku Lin}
\affiliation{Department of Physics \& Astronomy, University of
  California, Riverside, California 92521, USA}
  \affiliation{Department of Physics, Massachusetts Institute of Technology, Cambridge, MA 02139, USA}
\author{Paul van der Schoot}
\affiliation{Group Theory of Polymers and Soft Matter, Eindhoven University of Technology, P.O. Box 513, 5600 MB Eindhoven, The Netherlands}
\affiliation{Institute for Theoretical Physics, Utrecht University, Leuvenlaan 4, 3584 CE Utrecht, The Netherlands}
\author{Roya Zandi}
\affiliation{Department of Physics \& Astronomy, University of
  California, Riverside, California 92521, USA}

\date{\today}
\begin{abstract}
Electrostatic interaction is the driving force for the encapsulation by virus coat proteins of nanoparticles such as quantum dots, gold particles and magnetic beads for, e.g., imaging and therapeutic purposes. In recent experimental work, Daniel et al. [\textit{ACS Nano} \textbf{4} (2010), 3853-3860] found the encapsulation efficiency to sensitively depend on the interplay between the surface charge density of negatively charged gold nanoparticles and the number of positive charges on the RNA binding domains of the proteins. Surprisingly, these experiments reveal that despite the highly cooperative nature of the co-assembly at low pH, the \textit{efficiency} of encapsulation is a gradual function of their surface charge density. We present a simple all-or-nothing mass action law combined with an electrostatic interaction model to explain the experiments. We find quantitative agreement with experimental observations, supporting the existence of a natural statistical charge distribution between nanoparticles. \end{abstract}

\advance\textheight by 0.1in
\renewcommand{\dbltopfraction}{0.95}
\renewcommand{\topfraction}{0.9}
\renewcommand{\textfraction}{0.07}

\maketitle

\section{Introduction}
\textit{In vitro} reconstitution of single-stranded RNA viruses is driven by electrostatic interactions between positively charged RNA binding domains on the virus coat proteins also known as Arginine Rich Motifs or ARMs, and the negative charges on the backbone of RNA molecules. It is not entirely surprising then, that also non-native RNAs can be encapsulated by virus coat proteins, as synthetic polyanions, supramolecular polyanions and surface functionalized nanoparticles \cite{ZandiVanderSchoot2009,SiberPodgornik2009,SiberZandiPodgornik2010,VanderSchootBruinsma2005,SikkemaCornelissen2007,Shklovskii2008,RenWongLim2006,BelyiMuthukumar2006,HuGelbart2008,CadenaNavaGelbart2011,TingaWang2009,Hagan2009,MintenCornelissen2009,Bancroft1970,Bancroft1968,Bancroft1969,Hiebert1968}. Indeed, negatively charged gold, silica oxide, magnetite, as well as zinc and cadmium sulfide nanoparticles have been successfully encapsulated for the purpose of manufacturing meta materials and for imaging and therapeutic applications \cite{RenWongLim2006,CadenaNavaGelbart2011,ChenDragnea2006,YoungDouglas2008,HuangDragnea2007,DixitDragnea2006,JungRaoAnvari2011,Brasch,PotorskiSteinmetz2011,ShtykovaDragnea2007,Li2009,Saini2009,Farokhazad2004,Manchester2006}.

It is important to note in this context that the coat proteins of brome mosaic virus (BMV) and cowpea chlorotic mottle virus (CCMV) have been shown to spontaneously self-assemble \textit{in vitro} into \textit{empty} shells (capsids) under conditions of sufficiently low pH and high ionic strength in the absence of viral RNA \cite{Bancroft1970,Cuillel1987,Lavelle2009}. Under conditions of neutral or slightly basic pHs, virus-like particles form only in the presence of viral RNAs or other anionic cargos. Plausibly, electrostatic repulsion by the RNA binding domains, the ARMs, on the coat proteins prevents the self-assembly of empty shells, which would be of evolutionary advantage to the virus \cite{Bancroft1970,Bancroft1968,Bancroft1969,Hiebert1968,Cuillel1987,Lavelle2009,Verduin1969}. 

It turns out that not only the solution conditions are important but also the assembly protocol. This is illustrated in Figure 1, showing the reconstitution phase diagram of BMV from its constituent RNA and coat proteins as a function of pH and ionic strength obtained by Cuillel {\it et al.} by means of neutron scattering experiments \cite{Cuillel1987}. Starting off from the same initial conditions, indicated in the figure by a star, \textit{different} structures form depending on the assembly protocol followed.  If the pH is lowered first at high ionic strength then empty capsids form even in the presence of RNA, and filled ones are not found upon subsequent decrease of the ionic strength. However, if the ionic strength is lowered first and the pH is lowered after that, virus-like shells with RNA packaged inside them do form. It shows that once shells form, they do not easily disassemble and reassemble into a new structure even if thermodynamically this would be advantageous. Hysteresis effects observed in assembly and disassembly experiments confirm this \cite{VanderSchoot2009,Zlotnick2003,Verduin1974}.

\begin{figure}
\centering
\includegraphics[width=.4\textwidth,angle=90]{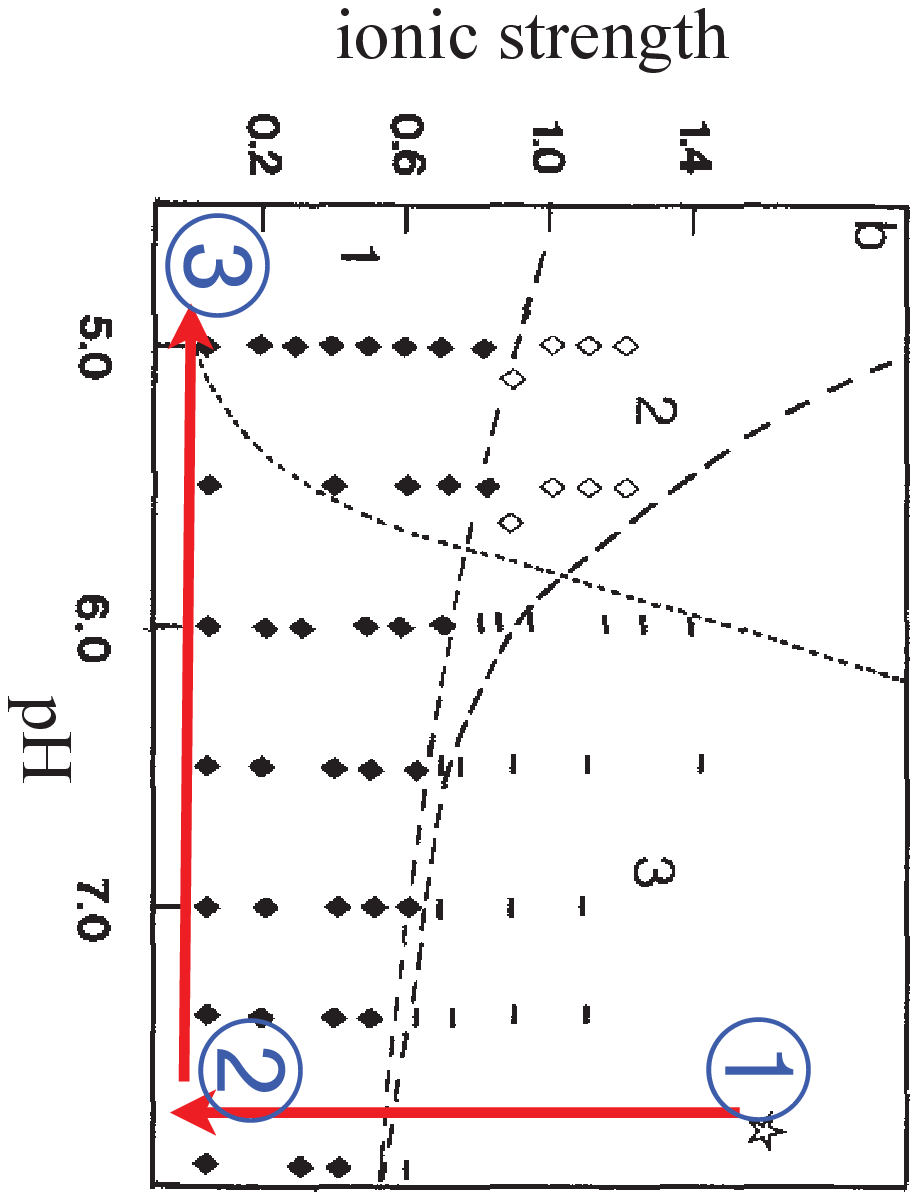}
\caption{Phase diagram for the \textit{in vitro} reconstitution of Brome Mosaic Virus BMV from viral RNA and coat protein, adapted from \cite{Cuillel1987}. The asterisk indicates the starting ionic strength and pH, and the arrows show the assembly protocol employed in Ref.~\cite{DanielDragnea2010} for the successful encapsulation of gold nanoparticles by BMV coat proteins. The symbols --, $\Diamond$, $\blacklozenge$ represent free coat proteins, empty capsids, and virus-like particles observed in solution respectively. All the measurements associated with the efficiency of the encapsulation of gold nanoparticles by BMV coat proteins discussed in the main text were performed at the point {\large \ding{174}}.}
\label{fig:reassembly}
\end{figure}

To obtain a better understanding of virus assembly in general and of the encapsulation of nanoparticles by virus coat proteins in particular relevant to many practical applications already alluded to, it is of interest to investigate the physical principles that determine the encapsulation efficiency and how this ties in with the assembly pathways. For this purpose, Daniel {\it et al.} have recently systematically investigated the impact of the surface charge density of functionalized gold nanoparticles on the formation of virus-like particles (VLPs) following the assembly protocol indicated in Fig.~\ref{fig:reassembly} by the thick line. The encapsulation experiments have revealed the following \cite{DanielDragnea2010,TsvetkovaDragnea2012,Chen2008,SunDragnea2007}:
\begin{itemize}
\item[1)] The size of the encapsulated nanoparticles directs the size of the virus-like particles albeit that their sizes do seem to correspond to the usual Caspar-Klug T numbers for viruses \cite{CasparKlug1962,Tnumber};
\item[2)] Nanoparticles only partially covered by viral coat proteins are rare at low pH. By far most particles are either completely covered or remain naked, indicating a high level of co-operativity of protein binding at low pHs;
\item[3)] The encapsulation efficiency, defined as the fraction of particles encapsulated to the total number of particles, increases monotonically with the surface charge density of the nanoparticles and with the ratio of protein to particle concentration;
\item[4)] A minimal negative surface charge density is required for encapsulation to be observed;
\item[5)] All other things remaining equal, the largest encapsulation efficiency appears to be obtained for nanoparticle sizes that their assemblies yield native virus capsid size;
\item[6)] Under conditions of low pH, co-assembly of nanoparticles and coat proteins is highly co-operative; yet, the efficiency of encapsulation is a gradual function of the surface charge density of the nanoparticles.
\end{itemize}

Theoretical studies by Hagan \cite{Hagan2009} and by \v{S}iber {\it et al.} \cite{SiberPodgornik2009,SiberZandiPodgornik2010} show that these findings can be rationalized within the context of the law of mass action, describing the co-assembly of proteins and nanoparticles, and Poisson-Boltzmann theory that describes the role of electrostatics in stabilizing the complexes. These and other works, including that of Prinsen {\it et al.} \cite{Prinsen2010} on the closely related problem of the stability of multishell capsids, show that surface charge density, nanoparticle size as well as the number of positive charges on and length of the ARMs dictate the encapsulation efficiency.

While qualitatively in agreement with experimental data, the cited theories cannot explain the combination of high co-operativity of the assembly and the rather gradual increase in encapsulation efficiency with increasing surface charge density \cite{DanielDragnea2010}.
In Ref. \cite{Hagan2009} the discrepancy between theory and experiment was explained in terms of metastable states and kinetic traps, i.e., issues relating to the assembly \textit{kinetics}. Here, we show that this discrepancy need not be of kinetic origin, but may be due to the existence of a broad distribution of surface charge densities of the nanoparticles associated with the way these surfaces are functionalized. Indeed, often nanoparticles are functionalized by attaching weakly acidic ligands to their surfaces, either for colloidal stability purposes, binding to assays or further biofunctionalization/compatibilization \cite{Rosenthal2011}.

To control the surface charge density, Dragnea and collaborators grafted short polymer molecules, polyethylene glycol or PEG, with charged and uncharged end groups onto the surface of gold nanoparticles \cite{ChenDragnea2006,DanielDragnea2010}. The charge density was controlled by varying the fraction of PEGs with acidic carboxyl end groups relative to those with neutral hydroxyl end groups. In fact, at fixed overall grafting density the surface charge density is determined not only by the fraction of end-grafted polymers with carboxyl end groups but also by the fraction of the weakly acidic carboxyl groups that are actually dissociated. For pHs where the assembly experiments were done, near the \textit{apparent} pK$_a$ value of the weakly acidic surface carboxyl groups \cite{Emoto1998, Latham2006}, we expect \textit{on average} about half the carboxyl groups to be dissociated. Note that the apparent pK$_a$ depends, in particular, on the size of the nanoparticles and the concentration of mobile ions \cite{Szleifer2011}.

It is important to realize that the statistical \textit{variation} of the fraction of ionizable groups around the average value is the largest for pHs around their pK$_a$ value. In macroscopic systems this is of little consequence but since we are dealing with nanoparticles, the number of carboxyl groups on them is not large and hence the variation around the mean must be important. Interestingly, accounting for this natural variability of surface charge density of the functionalized particles we find a very good match between theory and experiment, if we consider a Gaussian distribution around the average surface charge density of nanospheres and set the maximum charge density of nanoparticles to what is reported experimentally.

The mass action model that we invoke to describe the data is similar to that of earlier work, assuming (i) that the (effective) line tension associated with incomplete capsids is pronounced enough to prevent formation of significant fractions of intermediate states, and (ii) that the proteins do not form empty shells under the conditions that the experiments were done \cite{ZandiVanderSchoot2009}. Hence, we consider the nanoparticles can assume only two states, i.e., either completely covered or completely devoid of proteins, and calculate the ratio of encapsulated nanoparticles to the total number of nanoparticles as a function of their mean charge density.

For simplicity, we employ Debye-H\"{u}ckel theory to describe the electrostatic interactions between the various species, and for the width of the charge distribution of the nanoparticles we use a simple Langmuir adsorption model. Obviously, both non-linear Coulomb and charge regulation affect the numerical values in our calculations \cite{Szleifer2011}. However, they turn out not change the principle that we focus attention on, which is the natural fluctuation in charge density of nanoparticles \cite{kusters}. We find that while the stoichiometry ratio of nanoparticle and protein concentrations is an important factor that may make the all-or-nothing assembly seem less co-operative than it really is, within our equilibrium model the observed gradual variation of the encapsulation efficiency with increasing mean charge density of the nanoparticles can only be explained by the natural polydispersity of the charge density. Indeed, the \textit{size polydispersity} of the particles ($<10\%$) is too small to account for it \cite{TsvetkovaNote2012}.

It is important to emphasize that, as far as we know, the effect of charge variation has been previously ignored in the literature. The importance of this effect in the assembly of virus-like particles is due to the size of the nanospheres, which are much smaller than colloidal particles usually investigated. The existence of such a distribution is not all that significant for the usual colloidal particles that are tens to hundreds of nanometers in size. It does become important if the particles are small enough, i.e., are on the nanometer scale, and if charge regulation takes place. Since surface functionalization of nanoparticles with weakly acidic ligands is common, our findings should be of relevance not only to the encapsulation of gold particles, but also apply to quantum dots, magnetic beads and so on. In general, our findings should carry over to the electrostatic binding of nanoparticles onto macroscopic surfaces \cite{Fukuda2011}.

We note that for the purpose of this paper, we have considered that the solution of capsid proteins and cargo is in equilibrium and follow the law of mass action. Recently, self-assembly studies of Zlotnick have confirmed that the formation of a number of viruses follow a reversible path, suggesting that our assumption of equilibrium conditions is good. While the last step of capsid formation could be irreversible based on the studies of Zlotnick in Ref.~\cite{Zlotnick2007}, at the time scales relevant to the experiments, Zlotnick has shown that the concentration of proteins subunits and capsids approximately follow the law of mass action despite the possibility of irreversible steps towards the completion of full capsids. 

The remainder of this paper is organized as follows. In Section~\ref{sec:single_charge_model}, we present our ``two-state'' assembly model and derive the relevant formulas to calculate the efficiency of formation of virus-like particles in mixtures of nanoparticles and virus coat proteins as a function of various free binding energies and stoichiometry ratios. The effect of a fixed and a Gaussian surface charge distributions of the nanospheres is demonstrated in Sections~\ref{sec:fixed} and \ref{sec:variable}, respectively. Numerical results are illustrated in Section~\ref{sec:result}. We summarize our work and present our main conclusions in Section~\ref{sec:discussion}, where also discuss in more detail the justification for employing the approximate Debye-H\"{u}ckel theory. Details of our calculations are relegated to the appendix.

\section{Encapsulation of nanospheres}
\label{sec:single_charge_model}
As explained in the introduction, the experiments described in Ref. \cite{DanielDragnea2010} were performed under the conditions where virus-like particles do not form if their charge density is below a certain critical charge density--see item 4 in the previous section. This indicates that in these experiments, the concentration of protein subunits, $c$, must be below the critical subunit concentration, $c_*$, the minimum concentration necessary for formation of \textit{empty} capsids without polyanionic cargos. Otherwise, one would expect that the filled or empty capsids appear in solutions regardless of charge density of nanoparticles.  Further, the experiments reveal that the assembly process at pH 4.5 is highly cooperative: the nanoparticles are either naked or fully covered by protein subunits \cite{TsvetkovaDragnea2012,Chen2008}.

Based on these experimental observations, we consider a \textit{dilute} solution of capsid proteins with concentration c, and M negatively charged nanospheres. To mimic the experimental conditions, we assume that $c<c_*$ and thus there are no empty capsid in solution. Further, we consider that nanoparticles can assume only two states: either they are fully covered  with $q$ protein subunits or are completely naked. Let $\Delta g<0$ denote the binding free energy of a protein subunit adsorbed on a covered nanosphere. It accounts for the attractive hydrophobic interaction that drives the assembly in the absence of payload, the repulsive electrostatic interactions between proteins as well as the attractive electrostatic interaction between the proteins and the charged nanospheres.

The partition function of a semi-grand ensemble of coat proteins on the collection of $M$ empty and covered nanoparticles can then be written as

\begin{equation}
\mathcal{Z}=\sum_{N=0}^{M}\frac{M!}{N!(M-N)!}\lambda^N =(1+\lambda)^M
\end{equation}

\noindent with $\lambda \equiv \exp{[\beta(\mu-\Delta g)q]}$ the fugacity of the coat proteins on the nanoparticles, $\beta=1/k_BT$ the reciprocal thermal energy,  $k_B$ Boltzmann's constant and $T$ the absolute temperature. Furthermore, $q$ is the number of proteins on a virus-like particle (VLP) and $\mu$ the chemical potential of the proteins. The VLPS seem to obey the Caspar-Klug T-numbers, and thus we have $q=60 \times T$ \cite{CasparKlug1962,Tnumber}. Assuming the chemical equilibrium between the proteins in the VLPs and those in free solution, the chemical potentials must be equal.

The binomial distribution accounts for the number of indistinguishable ways of covering $N=0,1,...,M$ out of $M$ nanoparticles. How many of the $M$ particles are \textit{on average} covered by protein depends on the solution conditions through the chemical potential $\mu$ and binding free energy $\Delta g$. The efficiency $\eta$ defined as the ratio of  the number of nanospheres encapsulated by protein subunits to the total number of nanospheres is now straightforwardly calculated,
\begin{equation}
\eta\equiv\frac{\langle N \rangle}{M}=-\frac{\partial \ln \mathcal{Z}}{\partial \ln \lambda}=\frac{\lambda}{1+\lambda},
\label{eq:efficiency_definition}
\end{equation}
where $\langle N \rangle$ is the thermal average of the number of fully formed capsids. This is the familiar Langmuir adsorption isotherm.

In view of our discussion above, a quantity of particular interest is the strength of attractive interaction between the negatively charged nanospheres and protein subunits. It depends directly on the surface charge density of nanoparticles and drives the co-assembly. In the experiments of Ref. \cite{DanielDragnea2010}, the nanospheres (the gold particles) are coated with two different types of ligand: end carboxylated PEG (PEG--COOH) and end hydroxylated PEG (PEG--OH). The latter is neutral under all conditions of pH, while the former is a weak acid and hence can acquire a negative charge. For a given pH, the carboxyl group COOH can exist in two states of protonation: protonated and hence electrically neutral or deprotonated and negatively charged, depending on the acid constant or the negative logarithm of it, the $pK_a$. Hence, the surface charge density of nanospheres strongly depends on the relative numbers of the two ligands per unit area.

In the following section, we first assume that \textit{all} nanospheres have the same surface charge density that depends to some extent on the $pH$ of the solution. In the section following that, we consider the case in which the surface charge density varies among particles and obey a Gaussian distribution around an average value dictated again by the $pH$. The width of this distribution is calculated from a simple association-dissociation model.

\section{Fixed charge density}
\label{sec:fixed}
To calculate the efficiency of encapsulated nanoparticles, it is advantageous to rewrite Eq. \ref{eq:efficiency_definition} as a function of two relevant binding constants:
\begin{equation}
K=\exp{(-\beta \Delta g)}
\label{eq:binding_constant_K}
\end{equation}
and
\begin{equation}
K_0=\exp{(-\beta \Delta g_0)},
\end{equation}
where $\Delta g_0$ is the binding free energy associated with a subunit in an empty capsid. The latter describes the effect of attractive hydrophobic and repulsive electrostatic interactions between the coat proteins in an empty shells. We presume these interactions to be the same in empty and filled shells. Note that the electrostatic repulsion is dominated by the positive charges on the ARMs \cite{Kegel2004}.

The binding free energy difference between filled and empty shells can now be written as
\begin{equation}
q \Delta g-q \Delta g_0= \Delta f_E,
\end{equation}
which gives
\begin{equation}
\left(\frac{K}{K_0}\right)^q=\exp(- \beta \Delta f_E),
\label{eq:fe}
\end{equation}
\noindent with $f_E$ the electrostatic attractive interaction between a nanosphere and a protein subunit.

The attractive electrostatic interaction between nanoparticles and coat proteins is the result of negative charges on nanospheres and the positive charges on ARM regions of proteins. As explained in the introduction, the ARM region is a polypeptide chain that extends into the capsid interior and is rich in arginine and lysine, amino acids with strongly basic residues.

\begin{figure}
\centering
\includegraphics[width=.2\textwidth]{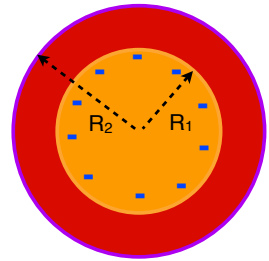}
\caption{Illustration of a virus-like-particle. The surface of inner shells is negatively charged. The N-terminal tails (the RNA binding domains) are
uniformly distributed in the darker shell and are positively charged. }
\label{fig:viral_shell}
\end{figure}

To calculate the electrostatic interaction between the protein shell and nanoparticles, we for simplicity assume that the charges on the ARMs are uniformly distributed in the space between the inner surface of the protein shell and nanoparticles, see Fig.~\ref{fig:viral_shell}. We then solve the relevant linearized Poisson-Boltzmann equation and find the electrostatic energy to obey
\begin{eqnarray}
&&\beta\Delta f_E\nonumber=\\
&&\frac{16 \pi^2  \lambda_B \sigma _1  \rho _2 R_1^2 \left( \left(\kappa  R_1+1\right)-e^{\kappa  (R_1-R_2)}
   \left(\kappa  R_2+1\right)\right)}{\kappa ^2 \left(\kappa  R_1 +1 \right)}\nonumber\\
\label{eq:electrostatic_interaction_energy}
\end{eqnarray}
with $\sigma_1<0$ the nanosphere surface charge density, $R_1$  the nanosphere radius plus the length of the ligands, $R_2$ the inner radius of capsid and $\rho_2>0$ its effective volume charge density due to the presence of the ARMs. The Bjerrum length, $\lambda_B=e^2/(4\pi \epsilon k_B T)$, where  $\epsilon$the dielectric permittivity of the medium, is $0.7$ $nm$ for water at room temperature. $\kappa = \sqrt{8\pi \lambda_D \rho_s}$ is the reciprocal of the Debye screening length with $\rho_s$ the number density of 1-1 electrolyte. At room temperature it is about $0.3/\sqrt{c_s}$, with $c_s$ the molar concentration of the electrolyte.  Equation \ref{eq:electrostatic_interaction_energy} is derived in Appendix~\ref{app:electrostatic_interaction_energy}.

Further headway can be made by noting that the reciprocal of the binding free energy $K_0$ is related to the critical capsid concentration, $c_* v = 1/K_0$, with $v$ the ``interaction'' volume, that we take to be the volume of a solvent molecule making $c_* v$ a mole fraction. The critical capsid density $c_*$ is a very important physical quantity that can be measured experimentally \cite{Zlotnick2002,Kegel2004}. Equation (\ref{eq:binding_constant_K}) can then be written as
\begin{equation}
\exp(\beta \Delta g)=\frac{K}{c_*vK_0},
\label{eq:binding_constant}
\end{equation}
where the quantity $K/K_0$ can be obtained from Eqs.~(\ref{eq:fe} and \ref{eq:electrostatic_interaction_energy}).

The expression for the efficiency, Eq. \ref{eq:efficiency_definition}, can be simplified further if we consider that the system is in chemical equilibrium and that the solution is dilute. The chemical potential $\mu$ of the free subunits is
\begin{equation}
\beta\mu=\ln c_f v,
\label{eq:chemical_potential}
\end{equation}
with $c_f$ the concentration of free subunits, which by mass conservation must be equal to
\begin{equation}
c_f=c-q\eta c_p=c(1-\eta r),
\label{eq:free_CPs}
\end{equation}
where $c$ is the overall concentration of protein subunits and $c_p$ that of the nanospheres in solution; the stoichiometry ratio $r$ is defined as $qc_p/c$ so that for a complete VLP $r=1$. Note that here we ignore a reference chemical potential. This merely renormalises $K_0$ and is implicit in the value of $c_*$.

Substituting Eqs.~(\ref{eq:binding_constant}),~(\ref{eq:chemical_potential}) and~(\ref{eq:free_CPs}) into Eq.~(\ref{eq:efficiency_definition}), we find the following expression for the efficiency $\eta$,
\begin{equation}
\eta=\frac{(\frac{c}{c^*})^q(\frac{K}{K_0})^q(1-\eta r)^q}{1+(\frac{c}{c^*})^q(\frac{K}{K_0})^q(1-\eta r)^q}
\label{eq:efficiency}
\end{equation}
This is our main result corresponding to the encapsulation efficiency of nanoparticles with a given surface charge density.

\begin{figure}[b]
\centering
\includegraphics[width=.4\textwidth]{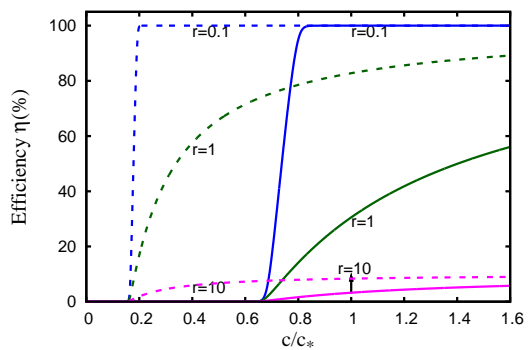}
\caption{The encapsulation efficiency $\eta$ versus the total protein concentration c, scaled to the critical capsid concentration $c_*$ for surface charge densities $\sigma_1 = 0.05$ (dashed lines) and $0.01$ (solid lines) $e/nm^{2}$, with three different values of the stoichiometric ratio $r=0.1,1,$ and $10$. For these plots, we have set the radius of the nanoparticles to $R_1=7.9$ $nm$, that of the cavity of the capsid to $R_2=11.4$ $nm$, $q=90$ (90 dimers assembling around the core) and the inverse Debye length $ \kappa=0.89$ nm$^{-1}$. We assume all PEG-COOH groups are deprotonated and negatively charged.}
\label{fig:etavscr}
\end{figure}

In the limiting case $K/K_0 \rightarrow 1$ and $c<c_*$, we find that the efficiency goes to zero, meaning that weak electrostatic interaction between the nanoparticles and protein subunits results into low efficiency of the formation of VLPs, as observed in the experiments \cite{DanielDragnea2010}. For large enough values of $K/K_0>c_*/c>1$, corresponding to strong attraction between nanospheres and coat protein, the efficiency goes toward unity as expected. In Fig.~\ref{fig:etavscr}, we plot the efficiency as a function of protein concentration scaled to the critical capsid concentration, $c/c_*$ for varying surface charge densities $\sigma_1$ and stoichiometry ratios r. Regardless of the values of $r$ the electrostatic interactions determine where the efficiency begins to rise from 0. As illustrated in the figure, the efficiency of VLPs gets smaller, as stoichiometry ratio becomes goes larger.
\section{Variable charge density}
\label{sec:variable}

To include in the theory the charge density distribution of nanoparticles, we explicitly consider the acid-base equilibrium of the carboxyl groups attached via the PEGs to the surface of the nanoparticles. The reaction scheme for a carboxylated PEG can be written as
\begin{center}
PEG--COOH
\xrightleftharpoons[\text{}]{\text{}}
PEG--COO$^-$+H$^+$
\end{center}
The (grand canonical) partition function $\mathcal{Z}_H$ associated with this equilibrium reads
\begin{equation}
\mathcal{Z}_H=1+\lambda_H
\label{eq:carboxylated_group_partition}
\end{equation}
with $\lambda_H=\exp(\beta \mu_H - \beta \epsilon_H)$ the fugacity of the hydrogen ions in solution, $\mu_H$  their chemical potential and $-\epsilon_H>0$ the free energy of ionization.  Note that because of chemical equilibrium, the chemical potential of the hydrogens on the PEG--COOHs and those in solution must be equal -- the ones in solution act as a reservoir.

In the partition function Eq.~(\ref{eq:carboxylated_group_partition}), the unity in the sum arises from the deprotonated state, PEG--COO$^-$, and the term $\lambda_H$ corresponds to the protonated state, PEG--COOH. The thermal average of the occupancy of H$^+$ binding sites PEG-COO$^{-1}$, forming PEG-COOHs, can be calculated from
\begin{equation}
\langle n \rangle=\frac{ \partial \ln Z_H}{\partial \ln \lambda_H}=\frac{\lambda_H }{1+\lambda_H},
\label{eq:thermal_average}
\end{equation}
which is equivalent to a Langmuir adsorption isotherm. The variance of the occupancy can be obtained from
\begin{equation}
\langle {\Delta n}^2 \rangle=\frac{\partial \langle n \rangle}{\partial \ln\lambda_H}=\frac{\lambda_H}{(1+\lambda_H)^2}.
\label{eq:variance_thermal_average}
\end{equation}

The fugacity $\lambda_H$ can be rewritten in terms of pH and pK$_a$ of the weak acid as follows \cite{Safran1994},
\begin{eqnarray}
\lambda_H=10^{\scriptstyle - pH + pK_a}
\label{eq:pH_pKa}
\end{eqnarray}
because, apart from a standard chemical potential that we incorporate in the ionization free energy $\epsilon_H$ , $\beta \mu_H=\ln (c_H v)$ with $c_H v$ the mole fraction of hydrogen ions in the solution. Note that $pH = - \log_{10} (55.6 \times c_H v)$ and $pK_a=-\log_{10} (55.6 \times \exp (-\beta \epsilon_H))$, where we used the molarity of water, $55.6$ $M$.
Substituting Eq.~(\ref{eq:pH_pKa}) into Eq.~(\ref{eq:thermal_average}), we obtain
\begin{equation}
\langle n \rangle=\frac{1}{1+10^{ \scriptstyle pH - pK_a}},
\end{equation}
with a similar expression for $\langle {\Delta n}^2 \rangle$. This is in essence the familiar Henderson-Hasselbalch equation.

We have to translate this expectation value for a single dissociable group to that for a collection of them on the surface of the nanoparticles. There, the effective $pK_a$ will depend on the strength of the electrostatic interaction between them, potential hydrogen bonding between neighboring COOH and COO$^-$ groups, and so on. In fact, the average protonation of the carboxyl groups on naked particles may well be different from that of encapsulated ones due to charge regulation \cite{Emoto1998,Latham2006}. This is an important issue but does not qualitatively change our
results. The focus of this paper is to show that statistical
fluctuations of the charged state of the particles has a large impact on
the perceived level of cooperativity of the encapsulation process.

For simplicity we, thus, assume that there exists a single pK$_a$ for both naked and covered particles that depends on the concentration of salt, the surface coverage of PEG--COOHs and some other factors. This implies that we ignore any coupling between the charged states of the carboxyl groups on the particles \cite{Szleifer2011,kusters}. Granting this simplification, which implies that our pK$_a$ is an effective one, the average surface charge density can be found by
\begin{eqnarray}
\langle\sigma\rangle=\sigma_{max}\frac{1}{1+10^{\scriptstyle -pH+pK_a}}
\label{eq:mean value}
\end{eqnarray}
with $\sigma_{max}$ the maximum surface charge density of the acidic groups that depends on the overall grafting density of PEGs and the ratio of the number of PEGs with COOH to those with OH end groups. This quantity is an experimentally known quantity \cite{DanielDragnea2010}.
Note that the PEG--OHs do not contribute to the charge density other than diluting the PEG--COOHs resulting into a decrease in $\sigma_{max}$.
The variance, then, is
\begin{equation}
\langle {\Delta\sigma}^2\rangle=\frac{\sigma_{max}^2}{q_\ell}\frac{10^{\scriptstyle -pH+pKa}}{(1+10^{\scriptstyle -pH+pKa})^2}
\label{eq:variance}
\end{equation}
where $q_\ell$ is the number of all the PEG-COOHs on the nanosphere. In the experiments of \cite{DanielDragnea2010}, the maximum value of $q_\ell$ is $\sim$3000 for a nanosphere with a diameter of 12 nm. In principle, we would like to know the full distribution $P(\sigma)$, rather than the first two moments of that distribution. However, because the large number of nanoparticles and the dilute solution, we invoke the central limit theorem assuming that the charge distribution is Gaussian. Hence, the probability density to find  a nanosphere with a surface charge density between $\sigma$ and $\sigma + \delta \sigma$, with $\delta \sigma$ an infinitesimal increment, is
\begin{equation}
\label{eq:normaldis}
P(\sigma)=\frac{1}{\sqrt{2\pi\Delta\sigma^2}}\exp\left\{{-\frac{(\sigma-\langle\sigma\rangle)^2}{2\Delta\sigma^2}}\right\}.
\end{equation}

The theory of the previous section for a fixed charge density needs to be modified to accommodate the charge distribution on the particles. To this end, we modify the conservation of mass, Eq.~\ref{eq:free_CPs},
\begin{equation}
c_f=c-qc_p\underbrace{\int_0^{\sigma_{max}} d\sigma P(\sigma)\eta(\sigma)}_{{\textstyle \eta}}
\label{eq:free_CPs_Gaussian}
\end{equation}
where $\eta(\sigma)$ is the efficiency of encapsulation of nano spheres for a given charge density $\sigma$ as described by Eq.~(\ref{eq:efficiency}).   The upper limit of the integral in Eq.~\ref{eq:free_CPs_Gaussian} is set to the maximum value of the surface charge density, $\sigma_{max}$.  The value of $\sigma_{max}$ is chosen assuming that all the -COOH groups are negatively charged, and is used in Eqs. \ref{eq:mean value} and \ref{eq:variance} to calculate the mean ($\langle \sigma\rangle$) and the variance ($\langle \Delta\sigma\rangle$) charge density. Due to small values of the variance ($\langle \Delta\sigma\rangle$), the portion of the Gaussian distribution corresponding to $\sigma<0$ is almost zero, and thus the lower limit of integral in Eq.~\ref{eq:free_CPs_Gaussian} is set to zero. Substituting Eq.~(\ref{eq:efficiency}) into Eq.~(\ref{eq:free_CPs_Gaussian}), we obtain
\begin{equation}
\frac{c_f}{c}= 1-r\int_0^{\sigma_{max}} d\sigma P(\sigma)\frac{(\frac{c}{c^*}\frac{K(\sigma)}{K_0(\sigma)}\frac{c_f}{c})^q}{1+(\frac{c}{c^*}\frac{K(\sigma)}{K_0(\sigma)}\frac{c_f}{c})^q}.
 \label{eq:efficiency_Gaussian}
 \end{equation}

The binding constants in Eq.~(\ref{eq:efficiency_Gaussian}) depend on the surface charge density of nanospheres. For a given $c/c_*<1$ and the stoichiometry ratio $r$, Eq.~(\ref{eq:efficiency_Gaussian}) can be solved  self-consistently and the ratio $c_f/c$ can be determined.  The total efficiency of encapsulation, $\eta$, is then obtained by the integral given in Eq.~(\ref{eq:efficiency_Gaussian}).

\begin{figure}[b]
\centering
\includegraphics[width=.4\textwidth]{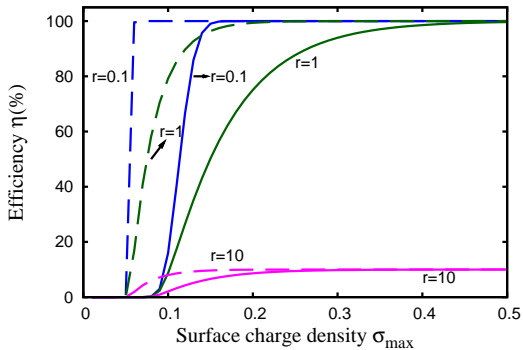}
\caption{The encapsulation efficiency $\eta$ versus the maximum surface charge density $\sigma_{max}$ $e/nm^{2}$ for two cases: pH=pK$_a$=4.6 (solid lines)and pH $\gg$ pK$_a=0$ (dashed lines) with three values of the stoichiometric ratio $r=0.1,1,10.$ The other model parameters are the same as given in Fig~\ref{fig:Gaussian_model}: $R_1=7.9$ nm, $R_2=11.4$ nm, $c/c_* = 0.14$.}
\label{fig:etavssigma}
\end{figure}

In Fig.~\ref{fig:etavssigma}, we plot the efficiency as a function of the surface charge density at pH = pK$_a$=4.6 and pH $\gg$ pK$_a$=0 for $r$ = 0.1, 1, and 10. The efficiency curves for pH $\gg$ pK$_a$=0 grow much faster and agree with the results for the fixed charge density model of the previous section, as expected.

\section{Comparison with experiments}
\label{sec:result}
As noted in the introduction, the efficiency of the encapsulation of functionalized nanoparticles by virus coat proteins was systematically measured as a function of their surface charge density by Dragnea and collaborators \cite{DanielDragnea2010}.  The experiments were performed at $pH=4.6$ and low ionic strength. The protein concentration and the assembly protocol, indicated by the thick line in Fig. 1, were chosen such that no \textit{empty} capsids were formed in the solution regardless of the charge density of gold nanoparticles. If the charge density of the nanoparticles was increased to a value above a critical charge density, capsids {\it filled} with nanospheres appeared in solution.  To describe these experimental observations, we employ the two-state model presented in the previous section.

According to the two-state model, the protein subunits are either free in solution or belong to VLPs, consistent with the experimental findings.  To compare our results with the experiments, we need to know the values of the following quantities: the inverse Debye length $\kappa$, the stoichiometric ratio $r$, the ratio of protein concentration to the critical concentration $c/c_*$, the number of protein subunits in each VLP, $q$, the radii $R_1$ and $R_2$, the effective volume charge density due to the presence of the ARMs $\rho_2$, the maximum surface charge density of the particles $\sigma_{\mbox max}$ and, finally, the effective $pK_a$ of the carboxyl groups and the $pH$ of the solution.

Unfortunately, there is not sufficient experimental data available to fix the values of all of these parameters. In particular, the values for $pK_a$ and $c/c_*$ are unknown for the conditions under which the experiments were done. Therefore, we use physically reasonable estimates to match our theoretical predictions with the experimental data associated with the assembly of gold nanoparticles by BMV capsid protein subunits \cite{DanielDragnea2010}. Note also that there is some freedom in the choice of $R_2$ too, because the inner surface of the capsid is not actually smooth but quite lumpy.

\begin{figure}[h!]
\centering
\includegraphics[width=.4\textwidth]{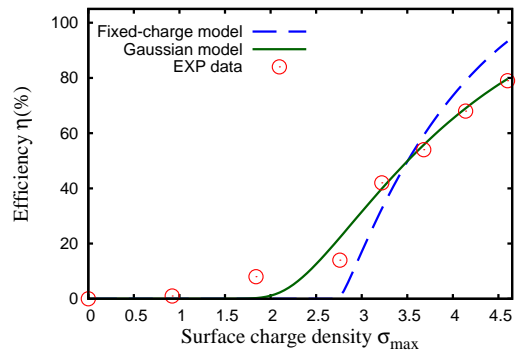}
\caption{The experimental data\cite{DanielDragnea2010} (red dots) compared to the Gaussian model (green solid line) and fixed charge density model (blue dashed line). The fitting parameters are $R_1=7.9$ nm, $R_2=11.4$ nm, $c/c_* = 0.14$, and pK$_a = 6.3$. The choice of the other model parameters is based on the experimental data: $q=90$ (90 dimers assembling around the gold nanoparticle), $r$ = 0.75 (the stoichiometry ratio), and $ \kappa=0.89$ nm$^{-1}$ (the inverse Debye length).}
\label{fig:Gaussian_model}
\end{figure}

\begin{figure}[h!]
\centering
\includegraphics[width=.4\textwidth]{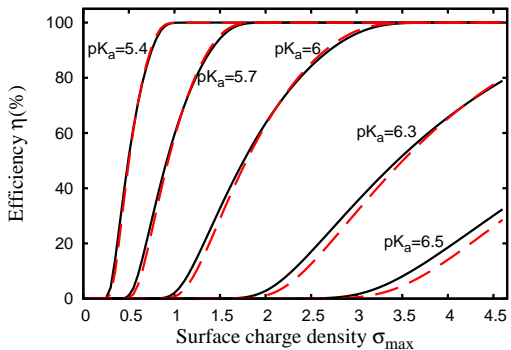}
\caption{ The fitting parameters for dashed lines are $R_1=7.9$ nm, $R_2=11.4$ nm, $c/c_* = 0.14$ and for solid lines are $R_1=7.2$ nm, $R_2=11$ nm, and $c/c_*=0.2$. The choice of the other model parameters is based on the experimental data: $q = 90$ (90 dimers assembling around the core), $r$ = 0.75 (the stoichiometry ratio), and $ \kappa=0.89$ nm$^{-1}$ (the inverse Debye length). The value of pK$_a$ is indicated next to each curve and varies from 5.4 to 6.5. }
\label{fig:pkavary}
\end{figure}

Figure~\ref{fig:Gaussian_model}  illustrates the efficiency vs. surface charge density obtained in the experiments (empty circles) for the radii $R_1=7.9$ $nm$ and $R_2=11.4$ $nm$, the stoichiometry ratio $r=0.75$, and $\kappa=0.89$ $nm^{-1}$. The dashed line in Fig.~\ref{fig:Gaussian_model}, corresponds to the efficiency obtained using the two-state model with the charge density of nanoparticles fixed. While the critical concentration, $c_*$, for the formation of empty capsids is not known for the $pH$ at which the experiments were done, the quantity $c_*/c$ associated with the experiments should be less than one, $c_*/c<1$, based on the fact that no capsids formed when the surface charge density of nanospheres was zero. Note that while the fit is not very sensitive to the small variations in $R_1$, $R_2$ and $c/c_*$, the value of pK$_a$ has a huge impact on the efficiency curves. This point is illustrated in Fig.~\ref{fig:pkavary} in which several efficiency curves for different values of pK$_a$ are plotted.  All dashed lines in the figures correspond to $R_1=7.9$ nm, $R_2=11.4$ nm, $c/c_* = 0.14$ and all solid lines correspond to $R_1=7.2$ nm, $R_2=11$ nm, $c/c_*=0.2$. The only variable among the solid or dashed curves is pK$_a$ as indicated in the figure.

As illustrated in fig.~\ref{fig:Gaussian_model}, the two-state model with a fixed charge density predicts the presence of a critical charge density below which no VLPs form, consistent with the experiments. However,  it does not capture one of the important features of the experimental data: The efficiency vs. charge density curves associated with the two-state model rise much faster than the experimental data.  To this end, we augment the two state model with a Gaussian charge distribution as explained in the previous section.

As shown in Eq.~\ref{eq:variance}, the width of Gaussian charge distribution depends on the dissociation constant (pK$_a$), which is not known for carboxyl group binding to PEG. Several experiments show that pK$_a$ depends on Coulomb and other types of interactions between neighboring carboxyl groups and also on the charge density of ARM regions.  To fit the data, we choose pK$_a = 6.3$, which is within the reasonable range of known values for the pK$_a$ of carboxyl end groups \cite{Emoto1998, Latham2006}. It is a bit below that of 11-mercaptoundecanoic
acid end grafted to gold nanoparticles under similar conditions of ionic strength \cite{Szleifer2011}. This we attribute to the effect of charge regulation resulting from the interaction of the carboxylic acid with the ARMs in the VLPs that we do not explicitly model in the present paper \cite{kusters}.

Despite the simplicity of our model, the solid lines in Figs. \ref{fig:Gaussian_model} show that the efficiency plots obtained with a Gaussian charge distribution gradually rise to unity rather than do so abruptly if we ignore the existence of a charge distribution. Indeed, there is a very good match between the theory and experimental data.

\section{Summary and Conclusion}
\label{sec:discussion}
Because the structure of transient assembly intermediates is not easily accessible experimentally, the underlying mechanisms of virus assembly are not, in general, very well understood. Still, several important theoretical and experimental works have recently shed light on the process of the formation \textit{in vitro} of virus capsids, in the presence and absence of polyanionic payload \cite{ZandiVanderSchoot2009,SiberPodgornik2009,SiberZandiPodgornik2010,VanderSchootBruinsma2005,SikkemaCornelissen2007,Shklovskii2008,RenWongLim2006,BelyiMuthukumar2006,HuGelbart2008,CadenaNavaGelbart2011,TingaWang2009,Hagan2009,MintenCornelissen2009, Zlotnick1994, Zlotnick2000, Zlotnick2003, Cuillel1987, Brasch}. Relevant to the work presented in this paper are the experimental studies of Dragnea et al. \cite{DanielDragnea2010,TsvetkovaDragnea2012} associated with the efficiency of incorporation of nano-payloads as a function of surface charge density. According to their studies, there is a critical charge density below which no encapsulated nanoparticles are observed.

A prominent feature of the self-assembly studies in Ref. \cite{DanielDragnea2010} is the dependence of the efficiency of incorporation of nanoparticles on surface charge density. Dragnea et al. found that beyond a critical charge density, the number of complexes of capsids and nanoparticles increases {\it gradually} as the charge density of nanoparticles increases. The {\it gradual} rise in the efficiency curves is unexpected because the process of capsid assembly  involves a high-order mass action that is inherently co-operative \cite{TsvetkovaDragnea2012}. About 180 protein subunits should assemble to form the T = 3 structure of BMV capsids, and this is the reason why there is always a very sharp rise in the the efficiency of encapsulation of nanoparticles beyond a critical protein concentration.

In order to understand the experimental results of Ref. \cite{DanielDragnea2010}, Hagan developed equilibrium and kinetic theories to explore the role of the charged nanoparticles in the assembly of virus-like particles \cite{Hagan2009}. His studies nicely show that there is indeed a threshold surface charge density above which nanoparticles become efficiently encapsuled by viral capsid proteins. However, the theoretical studies of Ref. \cite{Hagan2009} also show a sharp and very steep increase of the encapsulation efficiency with increasing charge density. Metastable states and the presence of kinetic traps were suggested as part of the reason for the discrepancy between the theory and experiments.

The central idea of this paper is that a natural, statistical charge variation between different nanoparticles, a  result of the way they are made to acquire charges not only in the above experiments but quite generally, must be the root cause of the discrepancy between previous theory \cite{Hagan2009} and the available experimental data  \cite{DanielDragnea2010}. The effect of charge variation has, as far as we know, been previously ignored in the literature. We argue that the importance of this effect in the assembly of virus-like particles is due to the size of the nanospheres, which are much smaller than colloidal particles usually investigated. The larger the particles are, the smaller the effect of charge variation is, which seems to be borne out by the experimental data presented in Ref.  \cite{DanielDragnea2010}.

We put forward a very simple model to describe the assembly of nano-cargos by virus coat proteins. In particular, we calculated the efficiency of nanospheres encapsulated by capsid proteins as a function of surface charge density, the coat protein concentrations, the stoichiometric ratio of protein to cargo concentrations, and interactions between protein and cargo. Our model is also able to explain successfully the presence of a critical charge density. We show that discrepancy between the theory presented in Ref. \cite{Hagan2009} and the experiments of Dragnea et al. can be explained by a distribution in the charge density of the nanoparticles.

Our work shows that the existence of a broad distribution of surface charges on nanoparticles stabilized by weakly acidic groups, can have a strong impact on their encapsulation by virus coat proteins. This is particularly true for pHs near the pKa of such groups, where the charge variation is most pronounced. The experiments cited were indeed performed close to the expected pKa between 5 and 7 for end-grafted polymeric surface layers bearing carboxyl moieties.

It is important to point out that our findings are not significantly affected by the Debye-H\"{u}ckel approximation that we employed in our calculations for reasons of simplicity.  Our goal was not to let complicated details of the theory requiring additional adjustable parameters cloud the issue. Indeed, including charge regulation and solving the non-linear Poisson-Boltzmann equation in the spirit of Prinsen and collaborators \cite{Prinsen2010}, we find that without charge variation the experimental data cannot be explained \cite{kusters}. As we alluded to in the introduction, also more elaborate theories \cite{Hagan2009} that in addition model the conformational statistics of the PEGs cannot explain the gradual rise of the encapsulation efficiency with surface charge density.

In view of the rapidly increasing number of self-assembly studies involving viral proteins and functionalized nanoparticles, we feel it is important to convey the message that there is an additional complication that is not considered: fluctuations become more important the smaller the particles. Clearly, comprehensive experimental and theoretical investigations of the parameters that affect capsid assembly in general, and the assembly of VLPs with poly- or macroanionic particles in particular, are highly desirable, and should have a significant impact on the development of optical imaging assays \cite{JungRaoAnvari2011}, therapies and treatment of viral infection \cite{Farokhazad2004}.

Finally we emphasize that the calculation done in this paper is quite general and could be applied to all nano-particles functionalized with weakly acidic groups.  Note that weakly acidic groups are commonly used in the context of biocompatibalizing nanoparticles \cite{ChenDragnea2006}.

The authors gratefully acknowledge helpful discussions with Mehran Kardar. We are grateful to Bogdan Dragnea and Irena Tsvetkova for discussions and providing the experimental data. This work was partially supported by the National Science Foundation through grant No. DMR-06-45668.

\appendix

\section{Electrostatic interaction energy}
\label{app:electrostatic_interaction_energy}
To calculate the electrostatic interaction between a nanosphere and N-terminal tail of capsid proteins, we consider an electrolyte solution with dielectric constant $\epsilon $. The electrostatic potential obeys the following linearized Poisson-Boltzmann equations in three different regions (see Fig. 2),
\begin{equation}
\nabla^2\phi=\left \{
 \begin{array}{ll}
 -\sigma_1\delta(r-R_1)/\epsilon,&\quad 0\leq r< R_1 \\
 \kappa^2 \phi-\rho_2/\epsilon,&\quad R_1\leq r < R_2 \\
 \kappa^2 \phi,&\quad r\geq R_2
 \end{array}\right.
\label{eq:ele}
\end{equation}
with the inverse of Debye length defined as $\kappa=\sqrt{2N_Ae^2 I/k_BT \epsilon}$, $I$ the ionic strength (M), $N_A$  Avogadro's number and $\epsilon$ the dielectric permittivity of the medium.

The general solution of eletrostatic potential can be found to be
 \begin{equation}
 \left \{
 \begin{array}{ll}
 \phi_1=C,&\quad 0\leq r< R_1 \\
 \phi_2=A \frac{\exp{(\kappa r)}}{r}+B \frac{\exp{(-\kappa r)}}{r}+\rho_2/\kappa^2\epsilon,&\quad R_1\leq r < R_2 \\
 \phi_3=D \frac{\exp{(-\kappa r)}}{r} ,&\quad r\geq R_2
 \end{array}\right.
\label{eq:ele}
\end{equation}
The four unknown parameters A, B, C, D can be obtained applying the boundary conditions at $R_1$ and $R_2$. The resulting electrostatic free energy can then be written as
\begin{equation}
f_E=\int d^3r \frac{\rho(r)\phi(r)}{2},
\end{equation}
which yields
\begin{equation}
f_E=\frac{4 \pi  e^2\sigma _1  \rho _2 R_1^2 \left( \left(\kappa  R_1+1\right)-e^{\kappa  (R_1-R_2)}
   \left(\kappa  R_2+1\right)\right)}{\epsilon\kappa ^2 \left(\kappa  R_1 +1 \right)},
   \label{eq:feapp}
\end{equation}
and self-energy to be
\begin{widetext}
\begin{equation}
f_{self}=\frac{2 \pi e^2\sigma _1^2 R_1^3 }{ \epsilon (\kappa  R_1+1) }-\frac{\pi  e^2\rho _2^2 \left(\left(\kappa  R_1+1\right) \left(2 \kappa ^3 R_1^3+\kappa ^2 R_2^2 \left(3-2 \kappa
   R_2\right)-3\right)-3 e^{2 \kappa  \left(R_1-R_2\right)} \left(\kappa  R_1-1\right) \left(\kappa  R_2+1\right){}^2\right)}{3
   \kappa ^5\epsilon \left(\kappa  R_1  +1 \right)}.
\end{equation}
\end{widetext}

Note that the contribution of the entropy of the ions is negligible within the Debye-H\"{u}ckel approximation. In Eq.~\ref{eq:feapp}, if $\sigma_1$ and $\rho_2$  have opposite signs then the binding constant $(K/K_0)^q=\exp({-\beta f_E})>>1$.

\end{document}